\documentclass[twocolumn,prl,showpacs]{revtex4}

\usepackage{amsmath}
\usepackage{amssymb}
\usepackage{epsfig}

\begin{document}
\title{Quantum Smoluchowski equation for driven systems}

\author{Raoul Dillenschneider}
\author{Eric Lutz}
\affiliation{Department of Physics, University of Augsburg, D-86135 Augsburg, 
Germany}

\date{\today}

\begin{abstract}
We consider a driven quantum harmonic oscillator strongly coupled to a heat bath. Starting from the exact quantum Langevin equation, we use a Green's function approach to determine the corresponding semiclassical equation for the Wigner phase space distribution.  In the limit of high friction, we apply Brinkman's method to  derive the quantum Smoluchowski equation for the probability distribution in position space. We further  determine the range  of validity of the equation  and discuss the special case of a Brownian parametric oscillator.
 
\end{abstract}


\maketitle

\emph{{Introduction.}}
The investigation of Brownian motion in the overdamped regime is of fundamental importance both from a theoretical and experimental point of view. In the limit of high friction, the velocity part of the phase space distribution of a Brownian particle quickly relaxes to equilibrium. Then for times much larger than the velocity relaxation time, a  description in terms of the position distribution  alone   becomes possible and the Kramers equation reduces to the simpler Smoluchowski equation \cite{risk89,cof96}.  Recently, a quantum generalization of the Smoluchowski equation for systems strongly coupled to a heat bath has been put forward \cite{Ankerhold,ank05}. The quantum Smoluchowski equation provides a semiclassical description of the evolution of the diagonal matrix elements of the system density operator in the coordinate representation; it allows to study the influence of thermal as well as  quantum fluctuations. The quantum Smoluchowski equation has been applied to problems involving both undriven and driven systems; examples  include the study of the quantum decay rates for driven potential barriers \cite{ank01}, quantum phase diffusion and charging effects in Josephson junctions \cite{ank04},   quantum diffusion in tilted periodic potentials \cite{mac06,cof08}, and  quantum extensions of nonequilibrium fluctuation theorems \cite{def09}. However, the exact expression of the quantum Smoluchowski equation for undriven systems is still the subject of discussions \cite{mac04,luc05,Coffey1,tse07,ank08} and a rigorous derivation for  driven quantum systems is lacking. 

The aim of this paper is twofold: We first present a transparent derivation of the quantum Smoluchowski equation that makes  use of controlled approximations, in order to complement the mathematically involved path integral derivation of Ref.~\cite{Ankerhold} and the heuristic approach of Ref.~\cite{Coffey1}. Second, we provide the first derivation of the driven quantum Smoluchowski equation  and determine its range of validity. In the following we consider a quantum harmonic oscillator with arbitrary time--dependent frequency. We employ a simple Green's function approach to determine the semiclassical expression of the diffusion coefficients appearing in the evolution equation of the Wigner phase space distribution, starting from the known exact quantum Langevin equation. In the large friction limit, we then apply the method introduced by  Brinkman \cite{Brinkman} to derive the quantum Smoluchowski equation. In the undriven case,  we recover the equation obtained in Refs.~\cite{Coffey1} and \cite{ank08}. In the driven case,  on the other hand, we show that the quantum Smoluchowski equation remains valid provided the driving rate is smaller than the velocity relaxation rate. We finally apply our results to the important example of the Brownian parametric quantum harmonic oscillator \cite{ZerbeHanggi}. 

\emph{{Quantum Langevin equation.}}
The starting point of our discussion is the quantum Langevin equation for a harmonic oscillator with time--dependent potential, $V(x,t) = m \omega^2(t)x^2/2$, linearly coupled to a bath of independent harmonic oscillators  \cite{for88,Ingold},
\begin{eqnarray}
m \ddot{x} + \int_0^t dt^{'} \gamma(t-t^{'}) \dot{x}(t^{'}) 
+\frac{\partial V}{\partial x} =
- \gamma(t) x_0 + F(t).
\label{Eq2}
\end{eqnarray}

\noindent
In the above equation the damping kernel is  given by $\gamma(t-t^{'}) = 1/\pi\,\int_{-\infty}^\infty d\nu \,
(J(\nu)/\nu) \cos  \nu (t-t^{'})$, where $J(\nu)$ is the spectral density of the bath. In the sequel we focus on the Ohmic regime where  $J(\nu) = m \gamma \nu$, the parameter $\gamma$  denoting the friction coefficient.
The fluctuating force operator $F(t)$  verifies the correlation function
\begin{eqnarray}
\langle F(t)F(t^{'}) + F(t^{'})F(t) \rangle = \hspace{3cm}\nonumber\\
\hbar \int_{-\infty}^{\infty} \frac{d\nu}{\pi} J(\nu)
\coth \left( \frac{\beta \hbar \nu}{2}\right)
\cos \nu (t-t^{'}) ,
\label{Eq3}
\end{eqnarray}

\noindent
where the average is taken over the initial bath degrees of freedom. Equation \eqref{Eq3} is exact and is derived under the assumption that the initial density operator of system plus bath factorizes. 
The solution of the Langevin equation \eqref{Eq2} can be written as,
\begin{eqnarray}
x(t) = m \dot{x}_0 G_1(t) + m x_0 G_2(t) + X(t),
\label{Eq4}
\end{eqnarray}

\noindent
where $x_0$ and $\dot{x}_0$ are the initial position and velocity of the quantum system and $X(t) = \int_0^t dt^{'} G(t,t^{'}) 
F(t^{'})$ is the fluctuating position operator. 
The two functions $G_1(t)$ and $G_2(t)$ are solutions of the homogeneous 
Langevin equation \eqref{Eq2} with vanishing member  on the right 
hand side.
By introducing  the two solutions,  $\phi_1(t) $ and $\phi_2(t) $, of the
equation $\ddot{y} + \Omega^2(t)y = 0$ with $\Omega^2(t) = \omega^2(t)
- \gamma^2 / 4$, we can explicitly write $G_1(t) = (1/m) \exp(-\gamma t /2) \,\phi_1(t)$
and $G_2(t) = (1/m) \exp(-\gamma t /2) \, \phi_2(t)$. 
The Green's function $G(t,t^{'})$ is then given by the
combination of the two functions $\phi_1$ and $\phi_2$ taken at different times
$t$ and $t^{'}$,
\begin{eqnarray}
G(t,t^{'}) = \frac{e^{\gamma (t-t^{'})/2}}{m}
\left(\phi_1(t) \phi_2(t^{'})
- \phi_1(t^{'}) \phi_2(t) \right).
\label{Eq5}
\end{eqnarray}

\noindent
A detailed derivation of the Green's function \eqref{Eq5} for the time--dependent harmonic oscillator  
can be found  in Ref.~\cite{Kleinert}.
The two functions $\phi_1(t)$ and $\phi_2(t)$ are linearly independent and their
Wronskian obeys $\dot{\phi}_1(t) \phi_2(t) - \phi_1(t) \dot{\phi}_2(t) =1$
for all time $t$. By fixing the initial conditions, $\phi_1(0) 
= \dot{\phi}_2(0) = 0$ and $\phi_2(0) 
= \dot{\phi}_1(0) = 1$, the Green's function $G(t,t^{'})$ verifies the relations $G(t,t) = 0$ and $G^{(1,0)}(t,t) = -G^{(0,1)}(t,t)
= 1/m$, where the numbers in parenthesis denote  derivatives with respect to the first (second) time argument. Moreover, since the system is initially decoupled from the heat bath, we have $\langle x_0 F(t) \rangle =  \langle \dot{x}_0 F(t)\rangle = 0$.

\emph{{Wigner function.}} The phase space dynamics of a quantum system is conveniently described using the Wigner quasiprobability distribution \cite{Wigner}. The evolution equation of the Wigner function $W(q,p,t)$ for a harmonic oscillator coupled to a bath of harmonic oscillators is of the general form \cite{sch85,HuPazZhang,HalliwellYu},
\begin{eqnarray}
\frac{\partial}{\partial t} W &=& 
- \frac{p}{m} \frac{\partial}{\partial q} W
+ m \widetilde{\Omega}^2(t) q
\frac{\partial}{\partial p} W
+ 2 \Gamma(t)
 \frac{\partial}{\partial p} (p W)
 \notag \\\
&&
+ D_{pp}(t)
 \frac{\partial^2}{\partial p^2} W
+ D_{qp}(t)
 \frac{\partial^2}{\partial q \partial p} W.
\label{Eq6}
\end{eqnarray}
An expression for the  time-dependent parameters $\widetilde{\Omega}^2(t)$, $2 \Gamma(t)$,
$D_{pp}(t)$ and $D_{qp}(t)$ for a driven harmonic oscillator can be directly derived from the quantum Langevin equation \eqref{Eq2} by generalizing the method introduced in Ref.~\cite{FordOConnell}.
The first step is to rewrite Eq.~\eqref{Eq2} in a form that is local in time \cite{han82}. This can be done by first inverting Eq.~\eqref{Eq4} and its  first time derivative
in order to express the initial coordinates $x_0$ 
and $\dot{x}_0$ in terms of the time-dependent variables. The latter coordinates are then injected into the second time derivative of  Eq.~\eqref{Eq4}, leading to a Langevin equation with time--dependent coefficients, 
\begin{eqnarray}
\ddot{x} + 2 \Gamma(t) \dot{x} + \widetilde{\Omega}^2(t) = F(t) / m.
\label{Eq7}
\end{eqnarray}

\noindent
Here we have defined the function
$2 \Gamma(t) = (G_1(t) \ddot{G}_2(t) - G_2(t) \ddot{G}_1(t))/
L(t)$ 
and the parameter
$\widetilde{\Omega}^2(t) = (\dot{G}_2(t) \ddot{G}_1(t)
- \dot{G}_1(t) \ddot{G}_2(t) ) / L(t)$. The  denominator is given by $L(t) = 
m ( G_2(t) \dot{G}_1(t) - G_1(t) \dot{G}_2(t))$. In the Ohmic regime, we simply have $2\Gamma(t) = \gamma$ and $\widetilde{\Omega}^2(t) = \omega^2(t)$. The second step is to derive an  equation similar to Eq.~\eqref{Eq7} from the evolution equation of the Wigner function \eqref{Eq6} by evaluating the first moments of the position and momentum operators,
$
\langle x \rangle = 
\int_{-\infty}^{\infty} dq 
\int_{-\infty}^{\infty} dp
(q + (i \hbar/2)  \partial_p) W(q,p,t)
$
and 
$\langle p \rangle = m\langle \dot x \rangle =
\int_{-\infty}^{\infty} dq 
\int_{-\infty}^{\infty} dp
(p - (i \hbar/2) \partial_q ) W(q,p,t)
$. This yields
\begin{eqnarray}
\langle \ddot{x} \rangle + 2 \Gamma(t)
\langle \dot{x} \rangle + \widetilde{\Omega}^2(t) \langle x \rangle =0
\label{Eq8}.
\end{eqnarray}

\noindent
Since the Langevin equation and the equation for the Wigner function describe the same process, the average of Eq.~\eqref{Eq7} must be equal to  Eq.~\eqref{Eq8}, implying that  the time-dependent 
parameters $2 \Gamma(t)$ and $\widetilde{\Omega}^2(t)$ are identical 
in both equations.
By repeating the same argument for the second moments of the position and momentum operators,
the diffusion coefficients $D_{pp}(t)$ and $D_{qp}(t)$ can be related  to the fluctuating position operator
$X(t)$ and the noise operator $F(t)$ via \cite{HalliwellYu,FordOConnell},
\begin{eqnarray}
D_{pp}(t) &=& \frac{m}{2} \langle \dot{X}(t) F(t) + F(t) \dot{X}(t)\rangle,\nonumber
\label{Eq9}
 \\
D_{qp}(t) &=& \frac{1}{2} \langle X(t) F(t) + F(t) X(t)\rangle.
\label{Eq10}
\end{eqnarray}

\emph{{Semiclassical diffusion coefficients.}}
The diffusion coefficients $D_{pp}(t)$ and $D_{qp}(t)$ can be further expressed
in terms of the correlation function \eqref{Eq3} of the noise operator and the Green's function \eqref{Eq5} by using the definition of the fluctuating position operator $X(t)$. We find,
\begin{eqnarray}
\label{10}
D_{pp}(t) &=& \frac{m}{2} \int_0^t dt^{'} G^{(1,0)}(t,t^{'}) 
\langle F(t^{'})F(t) + F(t)F(t^{'})\rangle, \nonumber
 \\
D_{qp}(t) &=& \frac{1}{2} \int_0^t dt^{'} G(t,t^{'}) 
\langle F(t^{'})F(t) + F(t)F(t^{'})\rangle.
\end{eqnarray}

\noindent
It is important to notice that the Green's function  $G(t,t^{'})$ is the same for quantum and classical oscillators and that the quantum--mechanical nature of the process is solely encoded in the noise correlation function. A semiclassical approximation of the diffusion coefficients can accordingly be obtained by expanding the noise correlator \eqref{Eq3} in powers of $\hbar$.  
 From Eq.~\eqref{Eq3}, we obtain  up to  second order for an Ohmic bath, 
\begin{eqnarray}
\langle F(t^{'})F(t) + F(t)F(t^{'})\rangle =
\frac{4 \gamma m}{\beta} \delta(t-t^{'}) \hspace{1cm}
\notag \\
- \hbar^2 \frac{\gamma m \beta}{3} \delta^{''}(t-t^{'})
+
\mathcal{O}(\hbar^3).
\label{Eq11}
\end{eqnarray}

\noindent
The first term on the right hand side corresponds to the classical noise correlation function, while the second term accounts for the first quantum corrections.
 We next evaluate the classical and quantum contributions to the  diffusion coefficients $D_{pp}(t)$ and $D_{qp}(t)$, Eqs.~\eqref{10}, separately.
The classical expression of 
the coefficient $D_{pp}(t)$ reads
\begin{equation}
\label{11}
D_{pp}^{c}(t) = \frac{m}{2} \int_0^t dt^{'} G^{(1,0)}(t,t^{'}) 
\frac{4 \gamma m}{\beta} \,\delta(t-t^{'})
= \frac{m \gamma}{\beta},
\end{equation}

\noindent
where we have used the relation $G^{(1,0)}(t,t) = 1/m$.
Since $G(t,t)=0$ the classical  
coefficient $D_{qp}(t)$  vanishes,
\begin{equation}
\label{12}
D_{qp}^{c}(t) = \frac{1}{2} \int_0^t dt^{'} G(t,t^{'}) 
\,
\frac{4 \gamma m}{\beta}\, \delta(t-t^{'})
= 0.
\end{equation}

\noindent
In the  limit $\hbar \rightarrow 0$ the diffusion coefficients are thus constant
and Eq.~\eqref{Eq6} reduces to the familiar  classical Klein--Kramers 
equation \cite{risk89,cof96}.

The quantum expressions of  the diffusion coefficients on the other hand
are obtained by  considering the quantum corrections to the noise correlation function  \eqref{Eq11}. 
For the coefficient $D_{pp}(t)$, we have,
\begin{eqnarray}
\label{13}
D_{pp}^{q}(t) &=& 
\frac{m}{2} \int_0^t dt^{'} G^{(1,0)}(t,t^{'}) 
\Big( 
- \frac{\gamma \hbar^2 m\beta}{3}\, \delta^{''}(t-t^{'})
\Big)
\notag \\
&=& \frac{2 \gamma m^2 \Lambda}{\beta} 
(
\omega^2(t) - \gamma^2
),
\end{eqnarray}

\noindent
where we have used  $G^{(1,2)}(t,t) = -(\omega^2(t) - \gamma^2)/m$
and  introduced the parameter $\Lambda = \hbar^2 \beta^2/24 m$ \cite{Coffey1}; the latter measures the magnitude of quantum fluctuations. By
noting moreover that $G^{(0,2)}(t,t) = - \gamma /m$, 
we find that the quantum contribution to $D_{qp}(t)$ is
\begin{equation}
\label{14}
D_{qp}^{q}(t) =
\frac{1}{2} \int_0^t dt^{'} G(t,t^{'}) 
\Big( 
- \frac{\gamma \hbar^2 m\beta}{3} \delta^{''}(t-t^{'})
\Big)
=
\frac{2 \gamma^2 m \Lambda}{\beta}.
\end{equation}

\noindent
By combining Eqs.~\eqref{11}, \eqref{12}, \eqref{13} and \eqref{14}, we eventually arrive at the following expressions for the semiclassical diffusion coefficients, up to second order in $\beta \hbar \omega$, 
\begin{eqnarray}
D_{pp}(t) &=& \frac{m \gamma}{\beta} 
+ \frac{2 \gamma m^2 \Lambda}{\beta} 
\left(
\omega^2(t) - \gamma^2
\right),
\label{Eq12}
 \\
D_{qp}(t) &=& \frac{2 \gamma^2 m \Lambda}{\beta}.
\label{Eq13}
\end{eqnarray}
Several points are worth emphasizing. First, for undriven systems, the diffusion coefficients are constant in time and Eq.~\eqref{Eq6} describes a Markovian process. This is due to the Ohmic regime and the semiclassical limit that we  consider here. Incidentally, in this regime the nature of the initial interaction between system and bath, decoupled or thermal, is not of importance. Furthermore, the quantum corrections $D_{pp}^{q}$ and $D_{qp}^{q}$ are proportional to the friction coefficient $\gamma$ and therefore become negligible  in the limit of vanishing coupling. We note in addition that  the terms $\gamma^2$ appearing in $D_{pp}^q(t)$ and $D_{qp}^q(t)$, Eqs.~\eqref{Eq12} and \eqref{Eq13}, are absent in the discussion of Ref.~\cite{Coffey1}. This difference can be traced back to the heuristic approach used in this work,  which consists in taking the limit  $\gamma \rightarrow 0$ for the evaluation of the diffusion coefficients. No such assumption is made here.

It is instructive to check the correctness of expressions \eqref{Eq12} and \eqref{Eq13} by computing the second moments of the position and momentum operators for an undriven harmonic oscillator,  $\omega^2(t) = \omega_0^2$. Starting from the equation for the Wigner function \eqref{Eq6}, one can show that \cite{Paz,MaiDhar},
\begin{eqnarray}
\langle x^2 \rangle &\underset{t \gg 1/\gamma}{=}& \frac{1}{\omega_0^2} \left( \frac{D_{pp}}{m^2 \gamma}
+ \frac{D_{qp}}{m} \right) =
\frac{1}{\beta m \omega_0^2} + \frac{2 \Lambda}{\beta}, \\
\langle p^2 \rangle &\underset{t \gg 1/\gamma}{=}&\frac{D_{pp}}{\gamma}=
\frac{m}{\beta} + \frac{2 m^2 \Lambda}{\beta} 
\left(\omega^2_0 - \gamma^2 \right),
\end{eqnarray}
which are in agreement with  the results obtained directly from the Langevin equation \eqref{Eq2}.

\emph{{Quantum Smoluchowski equation.}}
We are now in the position to derive the quantum Smoluchowski equation from the evolution equation of the Wigner function \eqref{Eq6}, in the limit of large friction. To this end, we use the method developed
by Brinkman \cite{Brinkman} and expand the Wigner function in the basis  of 
Weber functions $D_n(x)$ \cite{Zueco,Coffey1},
\begin{eqnarray}
W(q,p,t) = e^{-\beta \frac{p^2}{4m}}
\sum_{n=0}^\infty
D_n\left(p \sqrt{\frac{\beta}{2m}}\right) \varphi_n(q,t),
\label{Eq14}
\end{eqnarray}
\noindent
where $D_n(x) = 2^{-n/2} e^{-x^2 /2} H_n(x)$ and  
 $H_n(x)$ are the Hermite polynomials. 
The Weber functions are orthogonal, 
$\int_{-\infty}^\infty dy D_n(y) D_p(y) = (\pi n!) \delta_{n,p}$,
and obey the recurrence relations, 
$y D_n(y) = (n D_{n-1}(y) + D_{n+1}(y))/\sqrt{2}$
and $\partial_y D_n(y) = (n D_{n-1}(y) - D_{n+1}(y))/\sqrt{2}$.
By now inserting the expansion \eqref{Eq14} into Eq.~\eqref{Eq6} and integrating over the momentum variable after multiplying by 
$\sqrt{\beta/2m} \exp(\beta p^2 / 4m) D_n(p \sqrt{\beta/2m})$, 
one obtains the exact recurrence equation for the functions $\varphi_n (q,t)$,
\begin{eqnarray}
&&\partial_t \varphi_n 
+
\gamma n \varphi_n 
=
\notag \\
&&
-\frac{1}{\sqrt{\beta m}} 
\Big[
\partial_q \varphi_{n-1} + \left(n+1 \right) \partial_q \varphi_{n+1}
\Big]
-
\sqrt{\frac{\beta}{m}} \frac{\partial V}{\partial q} \varphi_{n-1}
\notag \\
&&
+ \left(
 \frac{\beta D_{pp} }{m}
-\gamma
\right)  \varphi_{n-2}
-D_{qp} \sqrt{\frac{\beta}{m}} 
\partial_q \varphi_{n-1}.
\label{Eq15}
\end{eqnarray}
\noindent
In the high damping limit, or equivalently in the noninertial limit $m \rightarrow 0$,  terms  of the form $\partial_t \varphi_n /\gamma$ in Eq.~\eqref{Eq15} become negligible for $n \ge 1$ and the Brinkman hierarchy can be simplified to,
\begin{eqnarray}
\frac{\partial \varphi_0}{\partial t} &=&
- \frac{1}{\sqrt{\beta m}} \frac{\partial \varphi_1}{\partial q}, \label{21}
 \\
\varphi_1 &=&
- \frac{1}{\gamma \sqrt{\beta m}} \Big(
\frac{\partial \varphi_0}{\partial q} 
+ 2 \frac{\partial \varphi_2}{\partial q}
+ \beta  V'(q,t) \varphi_0
\Big) 
\notag \\
&&
- 2 \sqrt{\frac{m}{\beta}} \gamma \Lambda 
\frac{\partial \varphi_0}{\partial q}, \label{22}
 \\
\varphi_2 &=&
- \frac{1}{2 \gamma \sqrt{\beta m}}
\Big(
\frac{\partial \varphi_1}{\partial q} 
+ 3 \frac{\partial \varphi_3}{\partial q}
+ \beta V'(q,t) \varphi_1
\Big)
\notag \\
&&
+ \left( \Lambda 
 V''(q,t)
- m \Lambda \gamma^2 \right) \varphi_0
-\sqrt{\frac{m}{\beta}} \gamma \Lambda 
\frac{\partial \varphi_1}{\partial q}.
\notag \\
\label{Eq16}
\end{eqnarray}
\noindent
In the limit $m \rightarrow 0$, the function $\varphi_2(q,t)$ further reduces to $\varphi_2(q,t) = \left(\Lambda  V''(q,t) 
- m \Lambda \gamma^2 \right)\varphi_0(q,t)$ and the set of three equations \eqref{21}, \eqref{22} and \eqref{Eq16} is   closed (the latter can be seen by multiplying Eq.~\eqref{Eq16} on both sides with $\sqrt{m}$ and by taking the $m$-dependence of  $\Lambda$ explicitly into account).  The probability density for the position of the Brownian particle is $P(q,t) = 
\int dp W(q,p,t) = \varphi_0(q,t)$. Solving Eqs.~\eqref{21}, \eqref{22} and \eqref{Eq16} for $\varphi_0(q,t)$, we then obtain the quantum Smoluchowski equation, 
\begin{eqnarray}
\frac{\partial P(q,t)}{\partial t} =\frac{1}{\gamma m}
\frac{\partial}{\partial q}
\left[  V'(q,t)
+\frac{1}{\beta}
\frac{\partial}{\partial q}
D_e(q,t) 
\right] P(q,t),
\label{Eq17}
\end{eqnarray}

\noindent
with $V'(q,t)= m \omega^2(t) q$ and the  diffusion coefficient $D_e(q,t) = ( 1 + 2 \Lambda V''(q,t))$. In the undriven case, Eq.~\eqref{Eq17}
 is equivalent to  the equation derived in Refs.~\cite{Coffey1,ank08} (in contrast to the one proposed in Refs.~\cite{Ankerhold,tse07}).
 Note that for thermodynamic reasons \cite{mac04,luc05}, the effective diffusion coefficient  $D_e(q,t)$ should be regarded 
as the first order expansion of   $D_e(q,t) = 1/( 1 - 2 \Lambda V''(q,t))$ (the argument presented in Refs.~\cite{mac04,luc05} only depends 
on the diffusion coefficient and not on the explicit form of the potential). The present derivation is restricted to harmonic potentials and a discussion of the quantum Smoluchowski equation for arbitrary  potentials will be given elsewhere.

The domain of validity of the quantum Smoluchowski equation \eqref{Eq17} can be determined from the condition 
$\partial_t \varphi_n/\gamma \ll \varphi_n$ \cite{Coffey1}. By introducing the characteristic length scale $\ell^2(t) = D \gamma / \omega^2(t)$ and using the replacement $\partial_x \varphi_1(q,t) \propto \varphi_1(\ell(t),t) / \ell(t)$, the requirement $\partial_t \varphi_1/\gamma \ll \varphi_1$ leads to $\omega(t)\ll \gamma$, $\hbar \omega(t) \ll kT$ and $\partial_t \omega(t) \ll \gamma^2$, which respectively correspond to  high--friction, high--temperature and moderate--driving conditions. The constraint $\partial_t \varphi_2/\gamma \ll \varphi_2$ further yields $\hbar \gamma\ll kT$. The condition $\partial_t \omega(t) \ll \gamma^2$ imposes that the driving rate is smaller than the velocity relaxation rate and ensures that the nondiagonal elements of the density operator of the driven quantum system remain negligible at all times.

\emph{{Parametric harmonic oscillator.}} 
An important example of a driven quantum system is the parametric oscillator with time--dependent frequency, $
\omega^2(t) = \omega^2_0 + \epsilon^2 \cos \left( \Omega_d t + \Phi \right)
$ \cite{ZerbeHanggi}. Here $\omega_0$ is the fixed   frequency of the oscillator, 
$\Omega_d$  the modulation frequency,  $\epsilon^2$  the
amplitude of modulation and $\Phi$  an initial phase. For this  exactly solvable system, the two functions $\phi_1(t)$ and $\phi_2(t)$ are given by Mathieu functions \cite{ZerbeHanggi}. We mention that the properties of a classical parametric oscillator have recently been investigated experimentally in optically trapped water droplets \cite{leo07}. For the case of the parametric oscillator, the  moderate--driving condition of the quantum Smoluchowski equation translates into $\epsilon \Omega_d \ll \gamma^2$, showing that for fixed friction coefficient, the restrictions on the driving frequency become more stringent, the larger the modulation amplitude.

\emph{{Conclusion.}}
We have presented a transparent and careful derivation of the quantum Smoluchowksi equation for a driven quantum system strongly coupled to a heat bath. Starting from the exact quantum  Langevin equation of a damped harmonic oscillator, we have combined a simple Green's function approach and a truncation of the Brinkman hierarchy in the strong friction limit to obtain the evolution equation of the semiclassical position distribution. Our findings confirm the results obtained  in Refs.~\cite{Coffey1,ank08} in the undriven case. On the other hand,  we have established the range of validity of the quantum Smoluchowski equation for a driven system and shown that it restricts the driving rate to be smaller than the velocity relaxation rate. We have finally discussed the important case of the parametric harmonic oscillator.

We thank S. Deffner, P. Talkner and P. H\"anggi for discussions. This work was supported by the Emmy Noether Program of the DFG 
(Contract LU1382/1-1) and the cluster of excellence Nanosystems Initiative 
Munich (NIM).


\end{document}